\documentclass[12pt,preprint]{aastex}
\usepackage{amsmath}

\shorttitle{Synchrotron Polarization of Thermal Electrons}

\shortauthors{Mao \& Wang}

\begin{document}


\title{Synchrotron Polarization of Relativistic Thermal Electrons}


\author{
Jirong Mao\altaffilmark{1,2,3} and
Jiancheng Wang\altaffilmark{1,2,3}
}
\altaffiltext{1}{Yunnan Observatories, Chinese Academy of Sciences, 650011 Kunming, Yunnan Province, China}
\altaffiltext{2}{Center for Astronomical Mega-Science, Chinese Academy of Sciences, 20A Datun Road, Chaoyang District, Beijing, 100012, China}
\altaffiltext{3}{Key Laboratory for the Structure and Evolution of Celestial Objects, Chinese Academy of Sciences, 650011 Kunming, China}

\email{jirongmao@mail.ynao.ac.cn}

\begin{abstract}
Relativistic electrons accelerated by both the first-order and the second-order Fermi accelerations in some synchrotron sources have a hybrid shape of thermal and nonthermal energy distribution. This particle acceleration result is supported by some recent numerical simulations. We calculate the synchrotron polarization by applying this electron energy distribution. The polarization degrees in the cases of active galactic nucleus (AGN) jets and gamma-ray bursts (GRBs) are given as examples. The possible application for the polarization study of Sgr $\rm{A^*}$ is also mentioned.
We finally suggest high-energy polarization measurements for these synchrotron sources to test our results.
\end{abstract}


\keywords{radiation mechanisms: non-thermal --- polarization --- gamma-ray burst: general --- BL Lacertae objects: general --- Galaxy: general}


\section{Introduction}
A relativistic electron has strongly polarized synchrotron emission when it moves in a large-scale and ordered magnetic field with a certain helical gyroradius orbit. The linear polarization degree is calculated by $p=K_{2/3}(x)/\int^{\infty}_{x}K_{5/3}(t)dt$, where $K(x)$ is the modified Bessel function, $x=\nu/\nu_c$, $\nu$ is the observational frequency, $\nu_c=\gamma^2\nu_L\rm{sin}(\alpha)$, $\nu_L=eB/m_ec$ is the gyrofrequency, $\alpha$ is the pitch angle, and $\gamma$ is the electron Lorentz factor. If the nonthermal relativistic electrons have a power-law energy distribution, which is usually produced by the first-order Fermi acceleration, the gross synchrotron emission has a linear polarization degree of $P=(3n+3)/(3n+7)$, where $n$ is the power-law index of the electron energy spectrum \citep{legg68,rybicki79}. This synchrotron polarization scenario has been widely applied for the study of synchrotron sources in the astrophysical field.

The general scenario mentioned above should be modified when we consider some complicated cases. Here, we present some motivations of our work. (1) In order to analytically obtain the liner polarization degree of $P=(3n+3)/(3n+7)$, the integrals of $xK_{2/3}(x)$ and $x\int K_{5/3}(x)dx$ can be mathematically derived to be the presentations with some Gamma functions. In this case, the lower and upper limits of the integrals are set to be zero and infinity, respectively \citep{legg68,rybicki79}. However, if the integrals are within a definite range, the analytical formulae of the above integrals that have the results simply related to the Gamma functions are not valid. In fact, the early work of Westfold (1959) presented one example to show the polarization degree that increases steadily with the frequency in the case that the electron energy spectrum has a given power-law shape. Therefore, the frequency-independent synchrotron polarization of $P=(3n+3)/(3n+7)$ is not universal. (2)
The electron energy distribution presented by a power law is usually produced by the first-order Fermi acceleration. However, the second-order Fermi acceleration can produce a Maxwellian electron energy distribution \citep{sch85,stawarz08}.
\citet{giannios09} proposed a mixture energy distribution of the power-law and the Maxwellian forms. This hybrid energy distribution of the relativistic electrons should be considered in the study of some synchrotron sources. These sources, such as active galactic nucleus (AGN) jets and gamma-ray bursts (GRBs), usually have strong relativistic shocks and turbulence. Thus, both the first- and second-order Fermi accelerations are important for the synchrotron polarization.
(3) Some numerical simulations of the particle acceleration showed that high-energy radiation can be generated by
fast-accelerated electrons with a hard spectrum in a small length scale \citep{cerutti14}. \citet{nal15} investigated the acceleration processes in detail under the framework of the plasmoid-dominated relativistic magnetic reconnection.  \citet{yuan16a} simulated the short-timescale polarization variabilities when the electron energy distribution has a Maxwellian shape with a power-law tail. The details of the plasmoid dynamics were illustrated by \citet{sironi16} and \citet{li17}. These simulation progresses encourage us to further explore the polarization processes. (4) There are some recent works related to the thermal electrons. \citet{res17} use the thermal electron energy distribution to derive the synchrotron self-absorption for the study of the GRB afterglow. The synchrotron radiation with the radiation transfer in the radio jets is presented when the Maxwellian distribution of the thermal electrons is given \citep{tso17}. However, the polarization properties are not provided in these works. (5) Some observational results of the synchrotron polarization have been accumulated. For example, \citet{steele17} recently summarized the optical polarization detections for nine GRBs. Because systematic analysis and explanations toward observational data samples are useful, a comprehensive study of the synchrotron polarization, including the effect from the relativistic thermal electrons, is necessary.

The synchrotron polarization contributed by the relativistic thermal electrons can be widely applied in the astrophysical research fields. We focus on the applications of AGN jets and GRB prompt emission in this paper. The polarization properties of Sgr $\rm{A^*}$ are also simply mentioned. The presentations of the hybrid electron energy distribution and the numerical calculations of synchrotron polarization are illustrated in Section 2. The polarization results corresponding to different AGN jet and GRB cases are presented in Section 3. We further discuss some possibilities to compare with observations and finally summarize in Section 3.

\section{Synchrotron Polarization of Thermal Electrons}
%
%
\subsection{Formulism}
The synchrotron polarization of a single electron is
\begin{equation}
p=\frac{K_{2/3}(x)}{\int_x^{\infty}K_{5/3}(t)dt},
\end{equation}
where $K(x)$ is the modified Bessel function, $x=\nu/\nu_c$, $\nu$ is the observational frequency, $\nu_c=\gamma^2\nu_L \rm{sin}(\alpha)$, $\nu_L=eB/m_ec$ is the gyroradius frequency, $\alpha$ is the pitch angle, and $\gamma$ is the electron Lorentz factor. In order to investigate the gross polarization of both thermal electrons and nonthermal electrons,
we adopt the mixture form of the electron energy distribution given by \citet{giannios09}. The distributions of the thermal and nonthermal electrons are
\begin{equation}
 N_{e1}(\gamma)=C_e\gamma^2\rm{exp}(-\gamma/\Theta)/2\Theta^3~~~for~~~\gamma \le \gamma_{th},
\end{equation}
and
\begin{equation}
N_{e2}(\gamma)=C_e\gamma_{th}^2\rm{exp}(-\gamma_{th}/\Theta)(\gamma/\gamma_{th})^{-n}/2\Theta^3~~~for~~~\gamma> \gamma_{th},
\end{equation}
respectively,
where $C_e$ is a constant, $n$ is the power-law index of the electron energy distribution, and $\gamma_{th}$ is the conjunctive Lorentz factor of electrons. Here, we take $\gamma_{th}$ as a variable parameter with the constraint of $\gamma_{min}<\gamma_{th}<\gamma_{max}$. The characteristic temperature is defined by $\Theta=kT_e/m_ec^2$, and $T_e$ is the thermal temperature of the relativistic electrons. We illustrate this hybrid electron energy distribution $N_e$ in the examples of $\gamma_{th}=3.0\times 10^3$ and $\gamma_{th}=5.5\times 10^4$ with different $T_e$ values in Figure 1.

In order to investigate the fraction of the electron energy that can be related to the thermal/nonthermal part of the
distribution, we follow the suggestion of \citet{giannios09} and propose one parameter $f$ that is defined as
\begin{equation}
f=\frac{\int_{\gamma_{th}}^{\infty}\gamma N_e(\gamma,\Theta)d\gamma}{\int_1^{\infty}\gamma N_e(\gamma, \Theta)d\gamma}.
\end{equation}
$N_e$ is determined by Equations (2) and (3). We present the parameter $f$ that indicates the energy fraction of nonthermal electrons as a function of the conjunctive Lorentz factor $\gamma_{th}$ in Figure 2. Clearly, $f=1$ in the condition of $\gamma_{th}=1$ means that we have an all non-thermal electron energy contribution, while thermal electron energy will be dominated by increasing $\gamma_{th}$.

The gross synchrotron polarization degree can be presented as
\begin{equation}
P=\frac{\int_0^{\pi/2}\int_{\gamma_{min}}^{\gamma_{th}}N_{e1}K_{2/3}(x)\rm{sin}(\alpha)d{\alpha}d\gamma+
\int_0^{\pi/2}\int_{\gamma_{th}}^{\gamma_{max}}N_{e2}K_{2/3}(x)sin(\alpha)d{\alpha}d\gamma}
{\int_0^{\pi/2}\int_{\gamma_{min}}^{\gamma_{th}}\int_x^{\infty}N_{e1}K_{5/3}(t)\rm{sin}(\alpha)dtd{\alpha}d\gamma
+\int_0^{\pi/2}\int_{\gamma_{th}}^{\gamma_{max}}\int_x^{\infty}N_{e2}K_{5/3}(t)\rm{sin}(\alpha)dtd{\alpha}d\gamma}.
\end{equation}
The synchrotron polarization degree mentioned above is strongly dependent on $\Theta$ and $\gamma_{th}$. We have $\Theta>1$ for the thermal electrons. If $\gamma_{th}$ approaches to $\gamma_{min}$, the polarization is dominated by the nonthermal electrons. On the other hand, the thermal electrons are dominated if $\gamma_{th}$ approaches to $\gamma_{max}$.

\subsection{Results in Different Cases}
The polarization results are strongly dependent on $\Theta$, which corresponds to the electron temperature. We take four temperature numbers ($T_e=10^{11}, 10^{12}, 10^{13}, \rm{and}~10^{14}$ K) of the thermal electrons to calculate the synchrotron linear polarization degree and apply the calculations to different cases. We put $\gamma_{min}=1.0$ and $\gamma_{max}=1.0\times 10^7$ in our calculations. The power-law index of the nonthermal electron distribution is fixed to be $n=2.5$.

The radio polarization measurements in the astrophysical fields are well established. We take the observational frequency of 5 GHz at C band.
We assume that the strength of the magnetic field in this case is 1 G. The polarization results with different numbers of $T_e=10^{11}, 10^{12}, \rm{and}~10^{13}$ K are shown in the top left panel of Figure 3. We suggest that this case can be adopted to study the radio polarization of AGN jets.

The polarization measurements in the optical band have also been widely performed. We set the observational frequency to be V band at $5500\AA$ and the strength of the magnetic field is 1 G in this case. The polarization results with different numbers of $T_e=10^{12}, 10^{13}, \rm{and}~10^{14}$ K are shown in the top right panel of Figure 3. We can use this case to study the optical polarization of AGN jets. This case may be also suitable to study the polarization of GRB optical afterglow, if the magnetic field in the GRB environment has the strength of about 1 G.

The polarization measurements in the X-ray band are interesting. We take the observational frequency of 5 keV in our calculations. The polarization results with different numbers of $T_e=10^{12}, 10^{13}, \rm{and}~10^{14}$ K are shown in the bottom left panel of Figure 3. Here, we change the magnetic field to be $1.0\times 10^4$ G. This case is proper for the polarization study of the early GRB X-ray afterglow. This case may be also suitable for the study of X-ray AGN jets, if we assume that AGN jets are strongly magnetized.

The gamma-ray polarization measurements have been proposed for some high-energy objects. We take the observational frequency of 5 MeV in our calculations. The polarization results with different numbers of $T_e=10^{12}, 10^{13}, \rm{and}~10^{14}$ K are shown in the bottom right panel of Figure 3. We set the magnetic field to be $1.0\times 10^6$ G. We suggest that this case can be used to study the polarization of the GRB prompt emission.

We further illustrate two issues in the four cases mentioned above. First, we confirm that the thermal relativistic electrons have effects on the synchrotron polarization. When the thermal electrons dominate the synchrotron radiation owing to the large conjunctive Lorentz factor, the linear polarization degree has a high value. In other words, the relativistic thermal electrons can produce high-degree polarized synchrotron radiation. On the other hand, we traditionally expect a highly polarized synchrotron radiation if all electrons are nonthermal and they have a power-law energy distribution. However, we obtain a relatively lower polarization degree when the nonthermal electrons dominate the radiation in this paper. We note that $\gamma_{th}$ splits electrons as thermal and
nonthermal parts. The thermal electrons have small Lorentz factor and the nonthermal electrons have large Lorentz
factor due to the electron distribution from Equations (2) \& (3). When we arbitrarily compare two independent cases, the case of all the thermal electrons with a Maxwellian distribution and the case of all the nonthermal electrons with a power-law distribution, the case of
nonthermal electrons has high polarization degree as well. However, a hybrid thermal-nonthermal electron energy distribution
seems more realistic in some astrophysical processes. We note that the electron energy distribution with the number of electrons is
dependent on the particle acceleration mechanisms in detail.
 We also understand from Equation (1) that the polarization degree of a single electron is a function of $\nu/\gamma^2\nu_L$. For a given frequency and a given magnetic field, one electron with a lower Lorentz factor has a higher polarization degree. Thus, in principle, the polarization dominated by the thermal electrons has a higher degree than that dominated by the nonthermal electrons.
Second, the transition range of the polarization degree between the thermal-electron-dominated phase and the nonthermal-electron-dominated phase is determined by the thermal electron temperature, because the electron energy distribution described by Equation (2) and (3) is regulated by $\Theta$. The transition process is determined by the term of $exp(-\gamma_{th}/\Theta)$.
When the thermal electrons have higher temperature, the polarization calculated by the term of $exp(-\gamma_{th}/\Theta)$ decreases slowly with increasing $\gamma_{th}$, and the transition of polarization as a function of $\gamma_{th}$ is smooth. When the thermal electrons have lower temperature, the number dominated by the term of $exp(-\gamma_{th}/\Theta)$ drops dramatically by an exponential way when we change $\gamma_{th}$, and the polarization has a very sharp transition.
Third, we show in Figure 3 that the thermal electrons with high
temperature have considerably lower polarization degrees. In fact, we see that the polarization degree with high electron
temperature corresponds to the case of a large $\gamma_{th}$ value. This means that the polarization degree with high electron
temperature is mainly contributed by the electrons with a large Lorentz factor.
As we mentioned above, one electron with a lower Lorentz factor has a higher polarization degree, while one electron with a higher
Lorentz factor has a lower polarization degree. Thus, we obtain the result that the thermal electrons
with high temperature produce relatively low polarization degrees.

In order to further explore the polarization degree properties related to the magnetic field
and the maximum electron Lorentz factor $\gamma_{max}$, we take one example in the case of $T_e=1.0\times 10^{13}$ K
at 5 MeV. Three $\gamma_{th}$ values, $3.0\times 10^3$, $2.0\times 10^4$, and $5.5\times 10^4$, corresponding to the
nonthermal-electron-dominated phase, nonthermal-thermal electron conjunctive phase, and thermal-electron dominated phase,
respectively, are selected. The results are shown in Figure 4. The polarization degree related to the magnetic field
is shown in the left panel. For both nonthermal- and thermal-electrons-dominated cases, the polarization degree slightly
decreases as the magnetic field increases. When we take the conjunctive case of thermal and nonthermal electrons,
the polarization degree increases as the magnetic field increases. The polarization degree related to the $\gamma_{max}$
is shown in the right panel. The polarization degree slightly decreases as we increase $\gamma_{max}$.
Because the range from $\gamma_{th}=5.5\times 10^4$ to $\gamma_{max}$ is narrow, such that the polarization degree has no
significant change with $\gamma_{max}$ when we take the thermal-electron-dominated case. We also note that $\gamma_{max}$
is determined by particle acceleration mechanisms in principle. We consider a thermal-nonthermal hybrid electron energy distribution in this paper. In principle, the electrons with the $\gamma$ values
from 1 to $10^7$ are all important to contribute the polarization. In particular, in our four cases presented in Figure 3,
both nonthermal electrons with large $\gamma$ values and thermal electrons with small $\gamma$ values have contributions to the
polarization in four different energy bands.

We pay attention that the usual polarization behavior of $p=(3n+3)/(3n+7)$ can be reproduced
with different parameters in our scenario. We get $p=(3n+3)/(3n+7)=72.4\%$ when $n=2.5$. As we see in Figure 3, for example,
we can reproduce this polarization degree at 5 MeV with the parameters of $T_e=10^{12}$ K, $B=1.0\times 10^6$ G, and $\gamma_{th}=7.9\times 10^3$.
We can also reproduce the same polarization degree at 5 keV with the parameters of $T_e=10^{12}$ K, $B=1.0\times 10^4$ G, and $\gamma_{th}=4.5\times 10^3$.
We note that the typical polarization degree given by the $(3n+3)/(3n+7)$ value is under
the condition of a purely nonthermal electron power-law energy distribution without any further constraint. However, when we consider
the electron energy distributions that described by Equations (2) and (3), it is clear that the synchrotron polarization is related to
$\gamma_{min}$, $\gamma_{th}$, $\gamma_{max}$, magnetic field, and $\Theta$ in each observational frequency. The polarization degree
in Equation (5) is strongly affected by the terms of $exp(-\gamma/\Theta)$ and $\nu/\nu_c$. Thus, the polarization degree in our study
deviates the $(3n+3)/(3n+7)$ value. Due to the contribution from the thermal component of the electron energy distribution and given that the
nonthermal electrons have a definite Lorentz factor range from $\gamma_{th}$ to $\gamma_{max}$ for the computation at a certain frequency,
the polarization produced by the nonthermal electrons does not simply reproduce the typical $(3n+3)/(3n+7)$ value. This deviation still
appears even if the nonthermal electrons dominate the synchrotron radiation. We present Figure 3 to illustrate four cases corresponding to
different astrophysical interests in detail. Both thermal and nonthermal electron effects of the synchrotron polarization are clearly
shown in each case due to the thermal-nonthermal hybrid electron energy distribution. Therefore, when we obtain a polarization degree of
about 72.4\%, we may have two possibilities at least. First, it can be produced by a pure nonthermal electron population with $n=2.5$.
Second, the polarization result may come from both thermal and nonthermal components of electrons. We suggest the second possibility in
our scenario.

The highly polarized photons produced by the relativistic thermal electrons can be detected in the radio, optical, X-ray, and gamma-ray bands. These results encourage the multi-wavelength polarization detections for the synchrotron sources. In particular, high-energy polarization observations are strongly suggested in this paper.




\section{Discussion and Conclusions}
In order to further constrain the particle acceleration mechanisms, we may compare the acceleration timescale and the cooling timescale for the thermal electrons. One simple estimation of the electron cooling timescale is
\begin{equation}
t_{cool}=\frac{6\pi m_ec}{\sigma_T\gamma_e B^2}=7.8\times 10^{-8}(\frac{\gamma}{1.0\times 10^4})^{-1}(\frac{B}{1.0\times 10^6~\rm{G}})^{-2}~\rm{s}.
\end{equation}
The large-scale stochastic acceleration that is thought to be one reasonable mechanism for the acceleration of the thermal electrons has been comprehensively investigated (e.g., Schlickeiser 1989a,b). \citet{petro04} calculated the timescales of the stochastic acceleration for both electrons and ions. Here, we take their results to estimate the electron acceleration timescale in the relativistic limit as
\begin{equation}
t_{ac}=\frac{q(q+2)\alpha^2}{4\gamma^{q-2}}\tau_p=1.7\times 10^{-7}(\frac{R}{1.0\times 10^9\rm{cm}})^2(\frac{n_e}{1.0\times 10^{10}\rm{cm^{-3}}})(\frac{f}{0.1})^{-1}(\frac{\gamma}{1.0\times 10^4})^{-1}(\frac{B}{1.0\times 10^6\rm{G}})^{-3}~\rm{s}.
\end{equation}
We take the index of the turbulent energy spectrum as $q=3$, and $\alpha=\omega_{pe}/\Omega_{e}$, where $\omega_{pe}$ is the plasma frequency and $\Omega_e$ is the electron gyrofrequency. The interaction timescale can be defined by $\tau_p^{-1}=(\pi/2)\Omega_e f(q-1)k^{q-1}_{min}$, where $f$ is the ratio between the turbulent energy density and the magnetic energy density, and we take $k_{min}=2\pi/R$ as the inverse of the acceleration size $R$. We choose the thickness of the GRB fireball that is defined by $R_{ball}/\Gamma^2$, where the fireball radius $R_{ball}\sim 10^{13}$ cm and the bulk Lorentz factor $\Gamma\sim 100$, to be the acceleration size $R$.

The radiation cooling timescale and the acceleration timescale are comparable if we consider the numbers of $B\sim 10^6$ G and $n_e\sim 10^{10}~\rm{cm^{-3}}$. This indicates that the stochastic acceleration could effectively generate the thermal electrons and the thermal electrons could provide the important contributions to the synchrotron polarization of the GRB prompt emission. If we assume the numbers of $B\sim 1$ G and $n_e\sim 1~\rm{cm^{-3}}$ for AGN jets, the radiation cooling timescale and the acceleration timescale are also comparable when we choose $R\sim 10^{11}$ cm. This means that we could also apply this work to explain the polarization in AGN jets if the turbulent eddies are very small compared to the large-scale jets.

We can also use this work to further discuss some short-timescale synchrotron polarization variabilities. For example, the optical polarization variabilities have been detected in some blazars during the strong gamma-ray flaring phase (e.g., Chandra et al. 2015). This short-timescale polarization variability was explained by the quick changing of the large-scale magnetic field \citep{deng16}. The optical polarization variability of GRB afterglow was also detected, and it could be interpreted by the topology variability of the large-scale magnetic field \citep{greiner03,gotz09,yonetoku11,wier12,mundell13}. In this work, we illustrate the large difference of the polarization degree varied with the conjunctive Lorentz factor. We speculate that the transition between the thermal-electron-dominated phase and the nonthermal-electron-dominated phase is popular in the dynamical evolution of GRB fireball and AGN jets. Therefore, we also expect the short-timescale variabilities of synchrotron polarization due to the transition of the electron energy distributions.

We focus on the relativistic regime in this paper. The condition of $\Theta=kT_e/m_ec^2>1$ should be satisfied.
This provides an electron temperature larger than $6.0\times 10^9$ K, while the $\gamma_{th}$ value is dependent on the different
mechanisms of the particle acceleration. In particular, if we consider different temperatures of electrons and ions,
the kinetic process that usually occurs in small length scale should be considered. Furthermore, the particle acceleration
in the magnetic-dominated plasmas has been investigated during recent years. Here, we take a recent example.
Nalewajko et al. (2016) have shown that the stochastic second-order Fermi process produces the electron power-law energy
distribution. Thus, $\gamma_{th}$ is determined by the turbulent acceleration in small length scales. However, there is no
any direct link between the electron temperature and the $\gamma_{th}$ value. In our paper, we illustrate that the
polarization degree changes with the $\gamma_{th}$ value. We expect more observations on this polarization behavior,
and we can constrain the particle acceleration processes by our polarization results.

The polarization results obtained in this paper are intrinsic. This means that we do not consider Faraday rotation, depolarization, and polarized radiation transfer processes. Because the high-energy polarized photons are effectively exempted from all kinds of depolarization effects, some X-ray and gamma-ray polarimeters, such as POLAR (Orsi 2010), ASTROSAT (Chattopadhyay et al. 2017), polSTAR \citep{kra16}, HARPO \citep{ber17}, eXTP \citep{zhang16}, IXPE \citep{wei16} and e-ASTROGAM \citep{tat17}, are strongly suggested for the polarization detection to the high-energy synchrotron sources.

Although we focus on the polarization study for GRBs and AGN jets in this paper, some other applications are also interesting for the synchrotron polarization of relativistic thermal electrons. For instance, the radio observations of the rapid polarization variabilities in Sgr $\rm{A^*}$ have been performed \citep{mar06,fish09,yusef11,huang12,liu16}, and the similar polarization variabilities have been also detected in the near-infrared band \citep{meyer06,eckart08,sha16}. If the stochastic acceleration is related to the flares of Sgr $\rm{A^*}$ \citep{liu04,liu06,chan09,petro12}, our modeling results can be further tested. Moreover, the X-ray flares of Sgr $\rm{A^*}$ have also been detected recently by NuSTAR, Swift and Chandra \citep{degenaar13,dibi14,ponti15,yuan16,zhang17}.
We also expect the X-ray polarization measurements for Sgr $\rm{A^*}$. Finally, we hope that
our model is helpful for the Event Horizon Telescope (EHT) project in the future \citep{chael16,cast17}.

\acknowledgments
We are grateful to the referee for his/her careful review and suggestions.
J. Mao is supported by the Hundred Talent Program, the Major Program of the Chinese Academy of Sciences (KJZD-EW-M06), the National Natural Science Foundation of China 11673062, and the Oversea Talent Program of Yunnan Province.
J. Wang is supported by the National Natural Science Foundation of China (11573060 and 11661161010).

\clearpage




\clearpage

\begin{figure}
\includegraphics[scale=0.3]{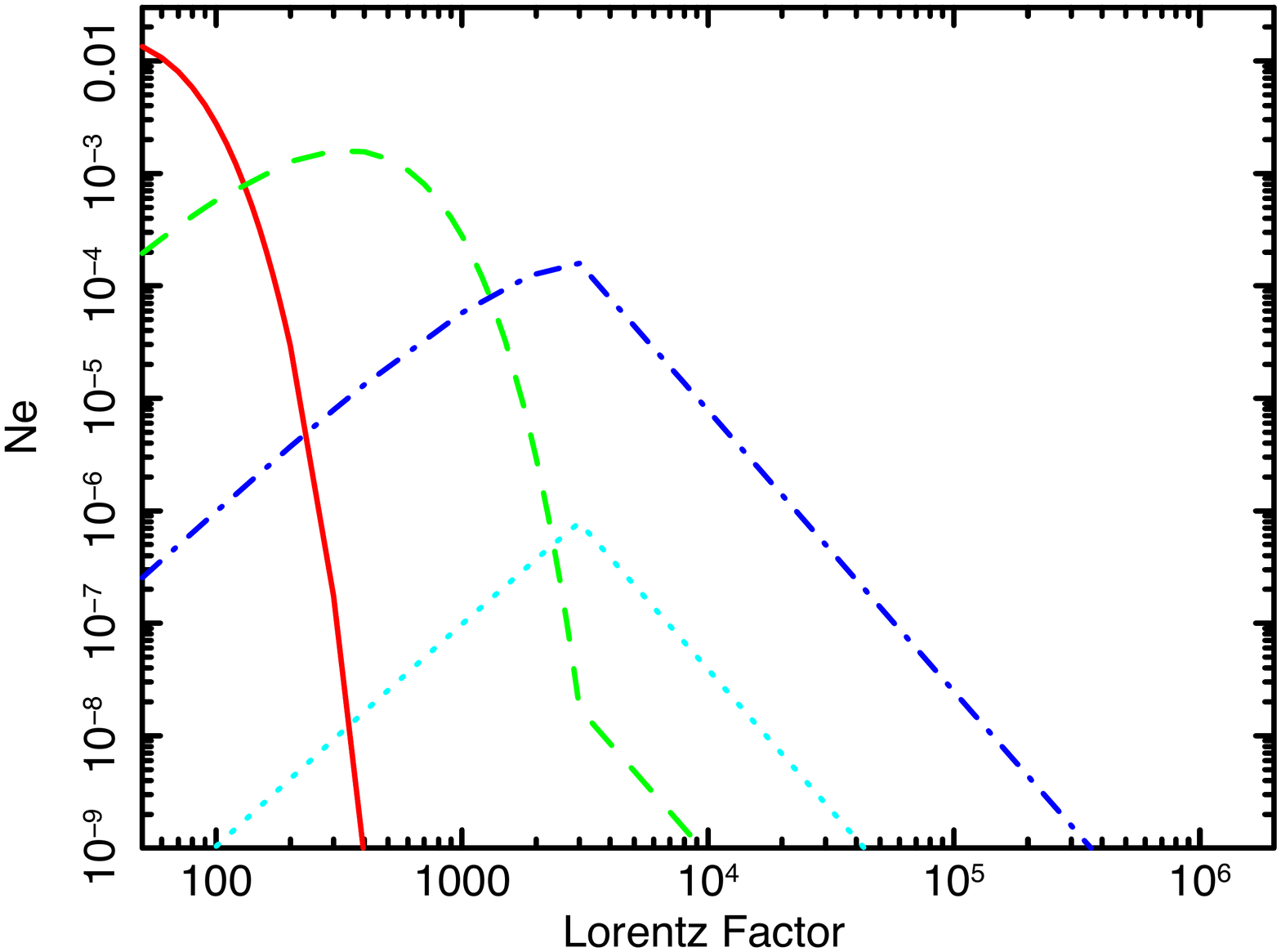}
\includegraphics[scale=0.3]{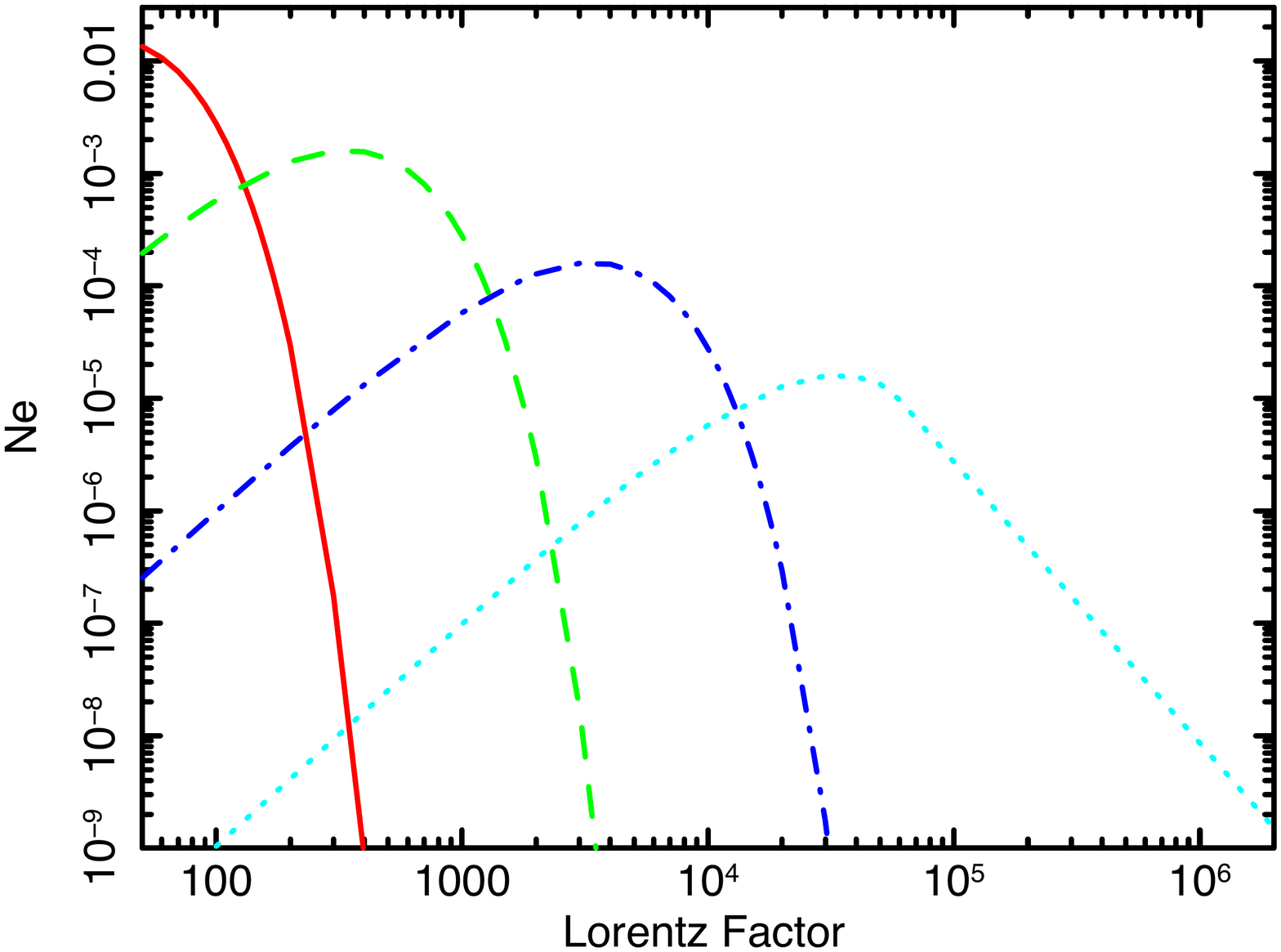}
\caption{Left panel: electron energy distribution $N_e$ with $\gamma_{th}=3.0\times 10^3$. Right panel: electron energy distribution $N_e$ with $\gamma_{th}=5.5\times 10^4$. The results are presented by the solid, dashed, dot-dashed, and dotted lines when we take $T_e=10^{11},~10^{12},~10^{13}~\rm{and}~10^{14}$ K, respectively.
\label{fig1}}
\end{figure}

\begin{figure}
\includegraphics[scale=0.6]{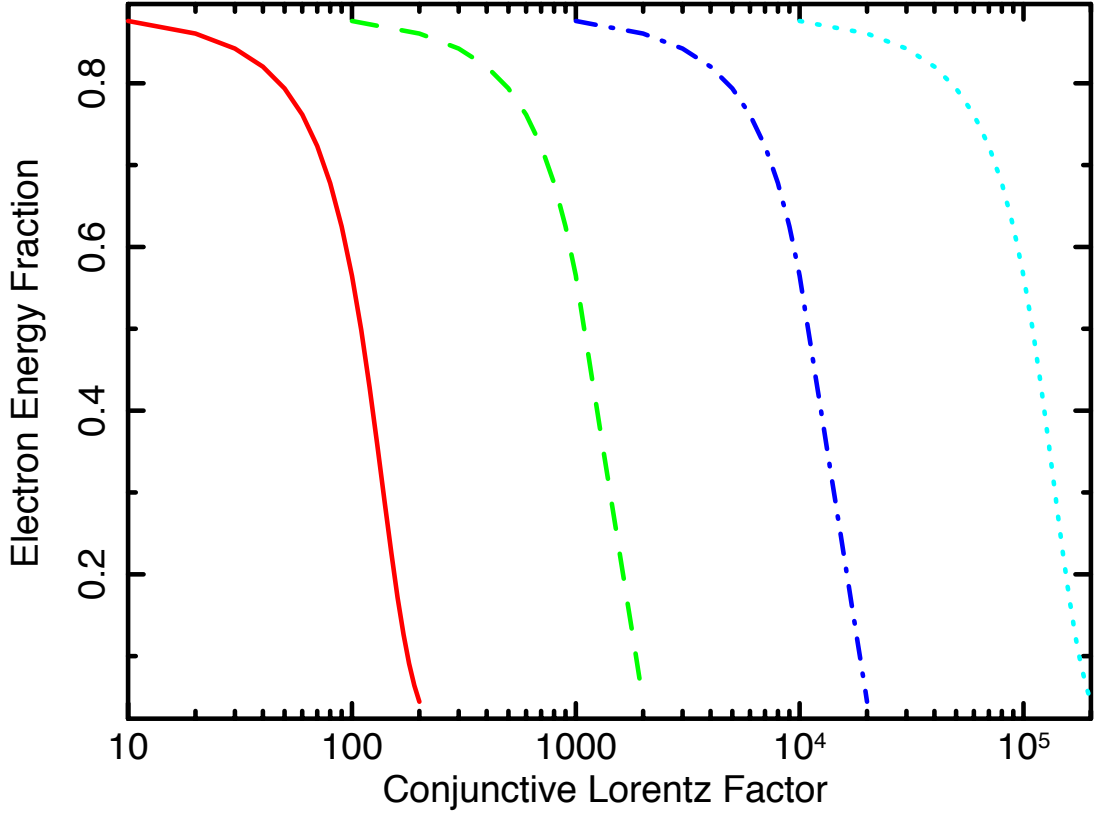}
\caption{Nonthermal electron energy fraction as a function of the electron conjunctive Lorentz factor $\gamma_{th}$. The results are presented by the solid, dashed, dot-dashed, and dotted lines when we take $T_e=10^{11},~10^{12},~10^{13}~\rm{and}~10^{14}$ K, respectively. In the calculation, we set the electron Lorentz factor within the range of $1<\gamma<10^7$.
\label{fig2}}
\end{figure}

\begin{figure}
\includegraphics[scale=0.3]{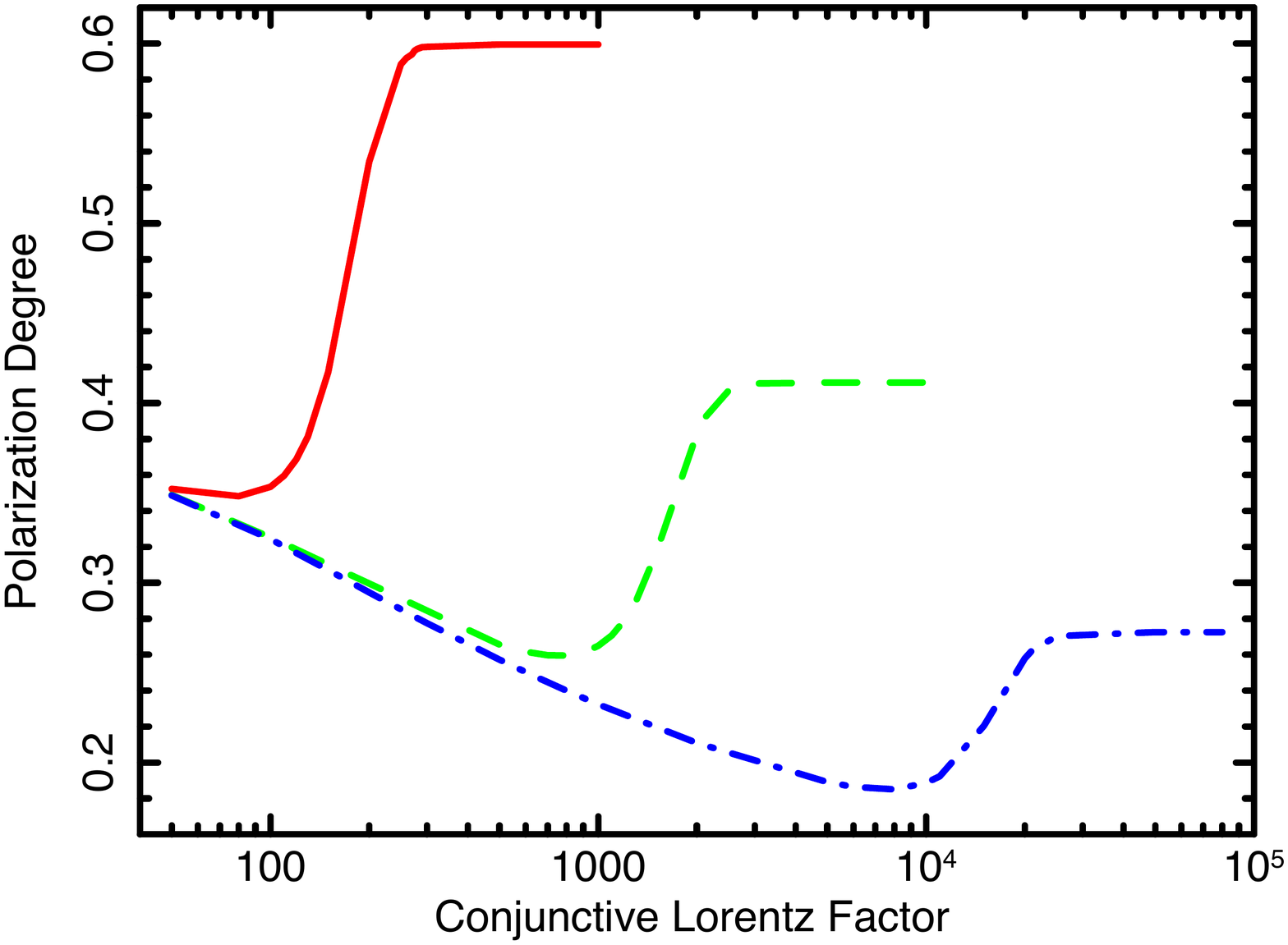}
\includegraphics[scale=0.3]{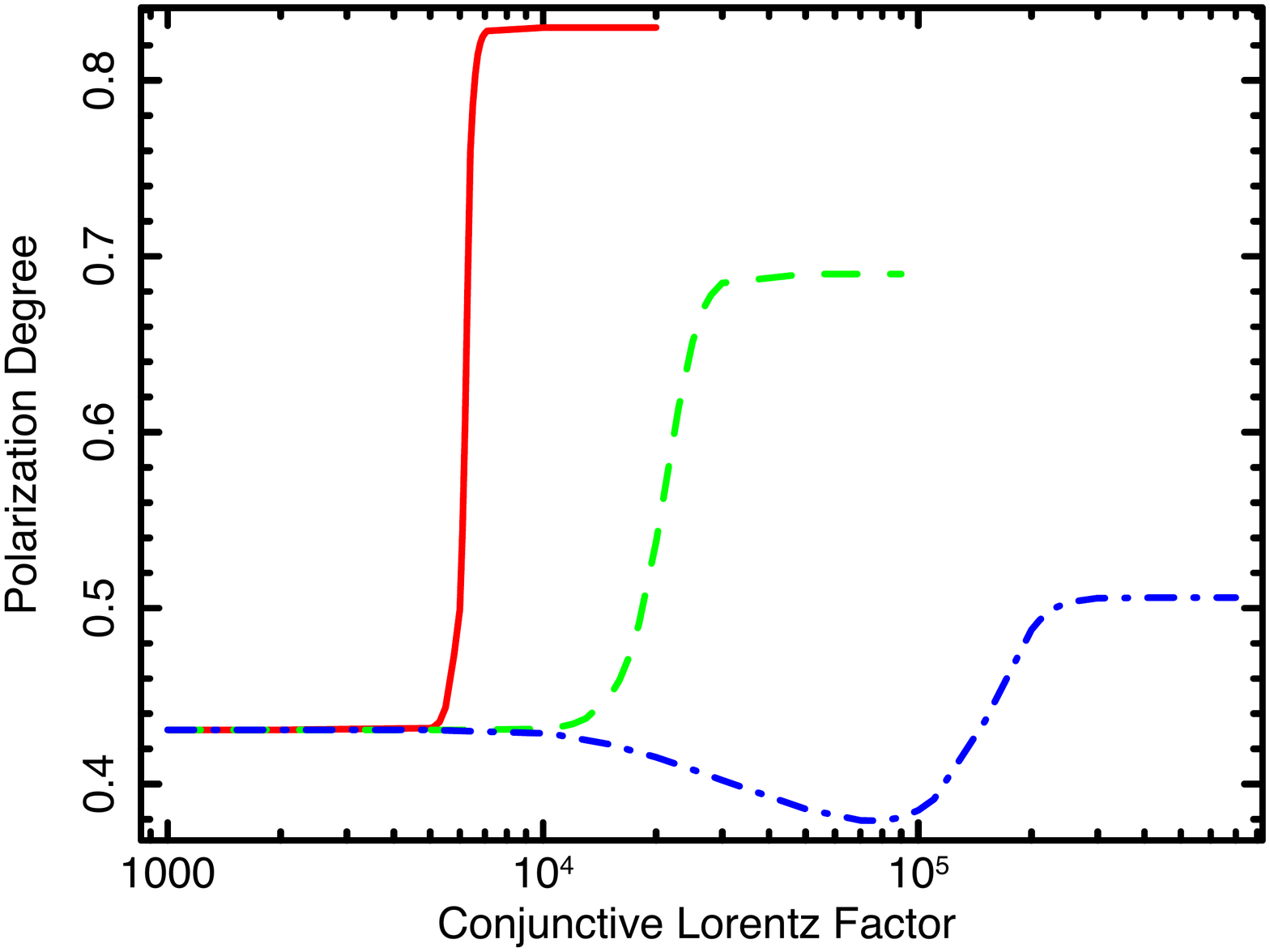}
\includegraphics[scale=0.3]{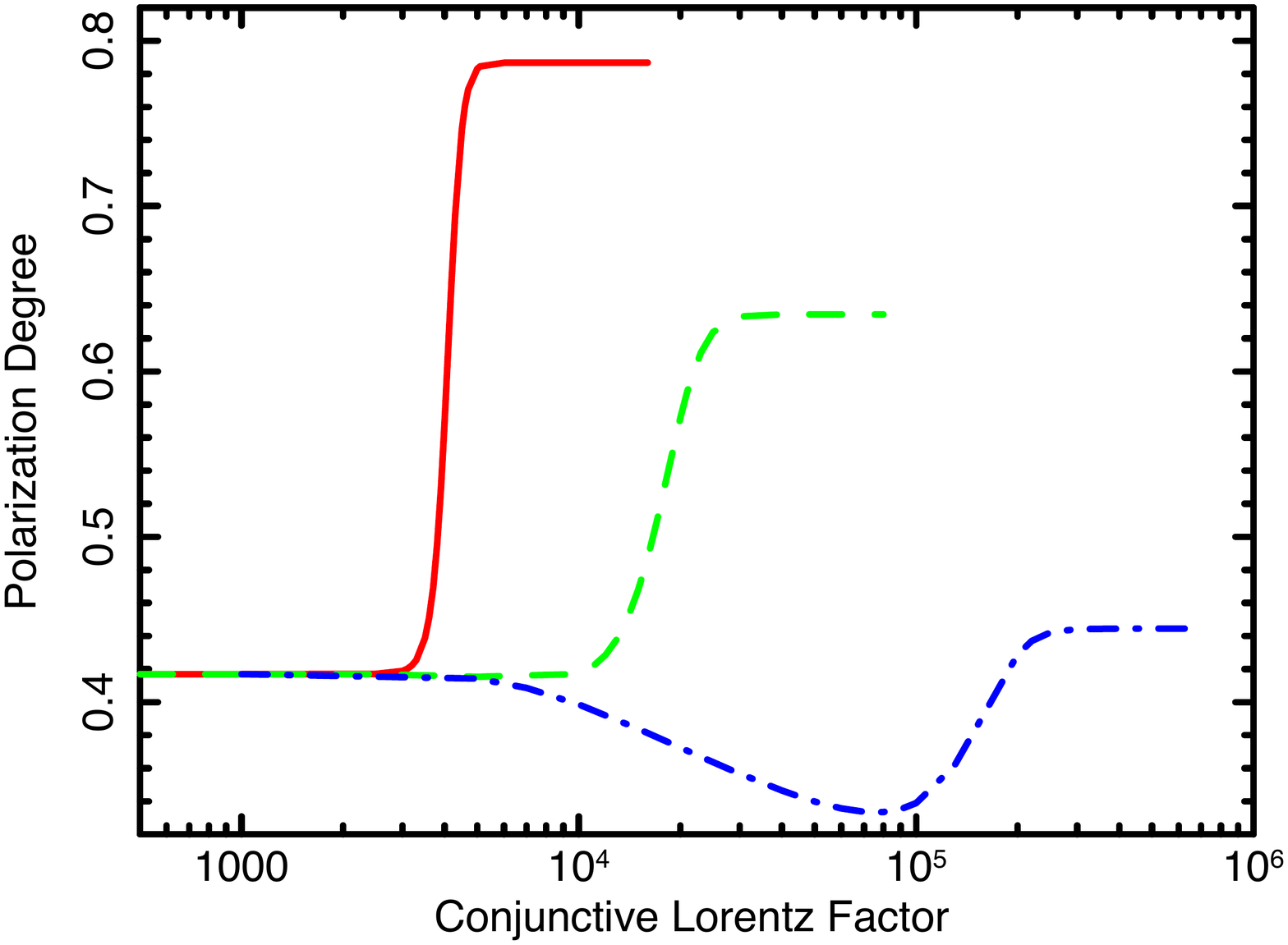}
\includegraphics[scale=0.3]{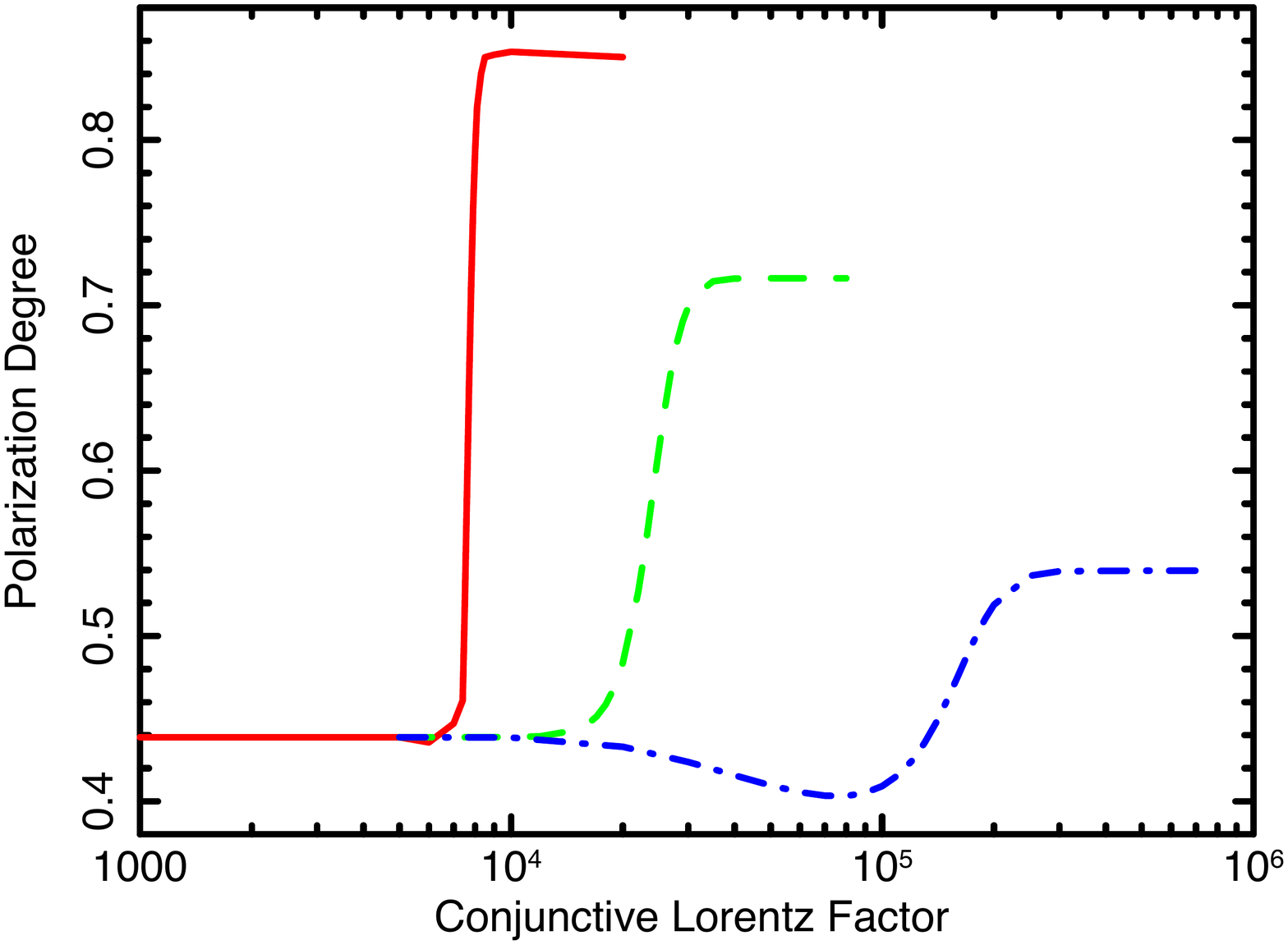}
\caption{Polarization degree as a function of the electron conjunctive Lorentz factor $\gamma_{th}$. The electron Lorentz factor has the range of $1<\gamma<10^7$. Top left panel: the polarization degree in the radio band (5 GHz) with $B=1$ G. The results are presented by the solid, dashed, and dot-dashed lines when we take $T_e=10^{11},~10^{12}~\rm{and}~10^{13}$ K, respectively. Top right panel: the polarization degree in the optical band (5500\AA) with $B=1$ G. The results are presented by the solid, dashed, and dot-dashed lines when we take $T_e=10^{12},~10^{13}~\rm{and}~10^{14}$ K, respectively. Bottom left panel: the polarization degree in the X-ray band (5 keV) with $B=10^4$ G. The results are presented by the solid, dashed, and dot-dashed lines when we take $T_e=10^{12},~10^{13}~\rm{and}~10^{14}$ K, respectively. Bottom right panel: the polarization degree in the gamma-ray band (5 MeV) with $B=10^6$ G. The results are presented by the solid, dashed, and dot-dashed lines when we take $T_e=10^{12},~10^{13}~\rm{and}~10^{14}$ K, respectively.
\label{fig3}}
\end{figure}

\clearpage
\begin{figure}
\includegraphics[scale=0.3]{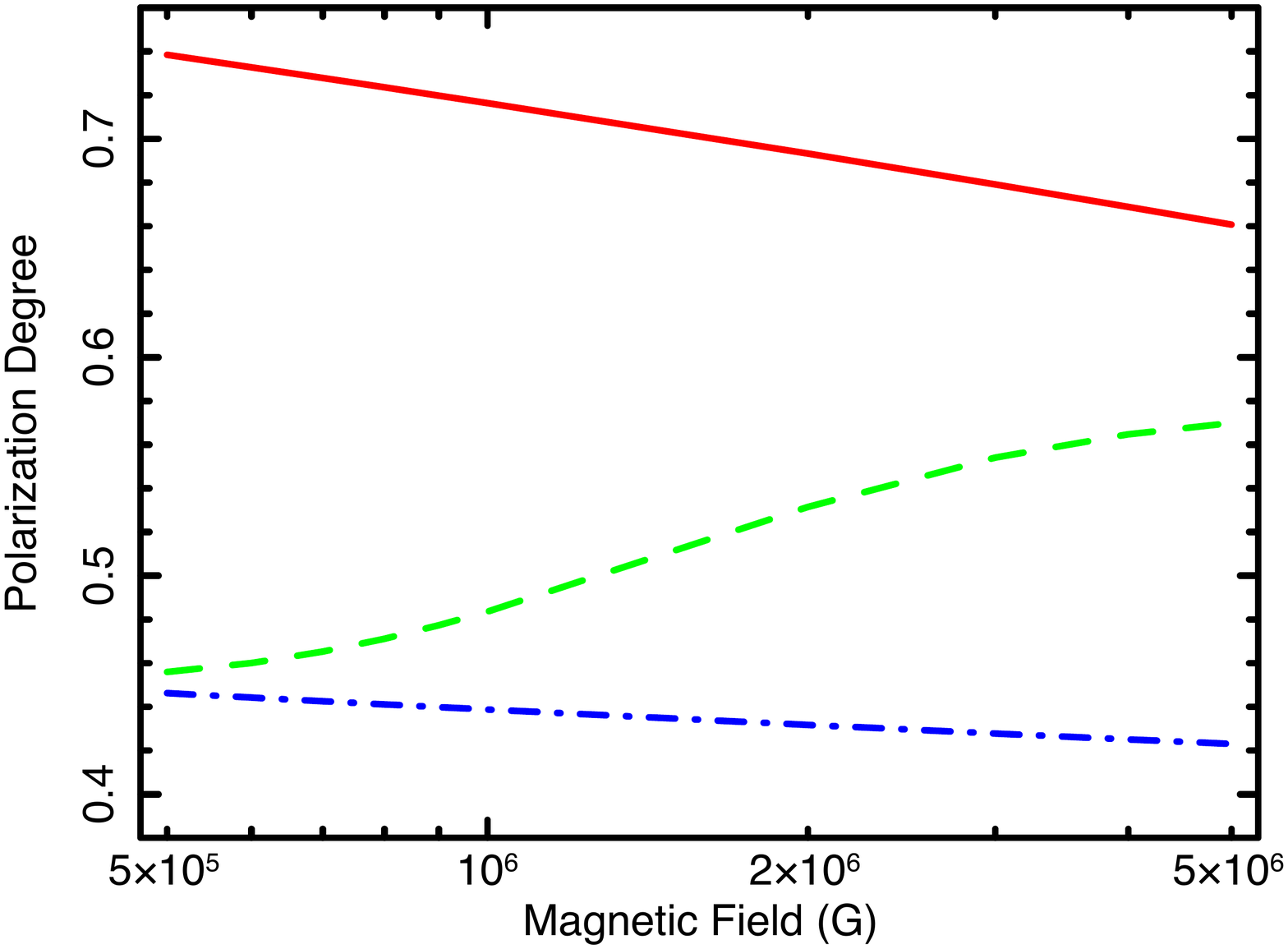}
\includegraphics[scale=0.3]{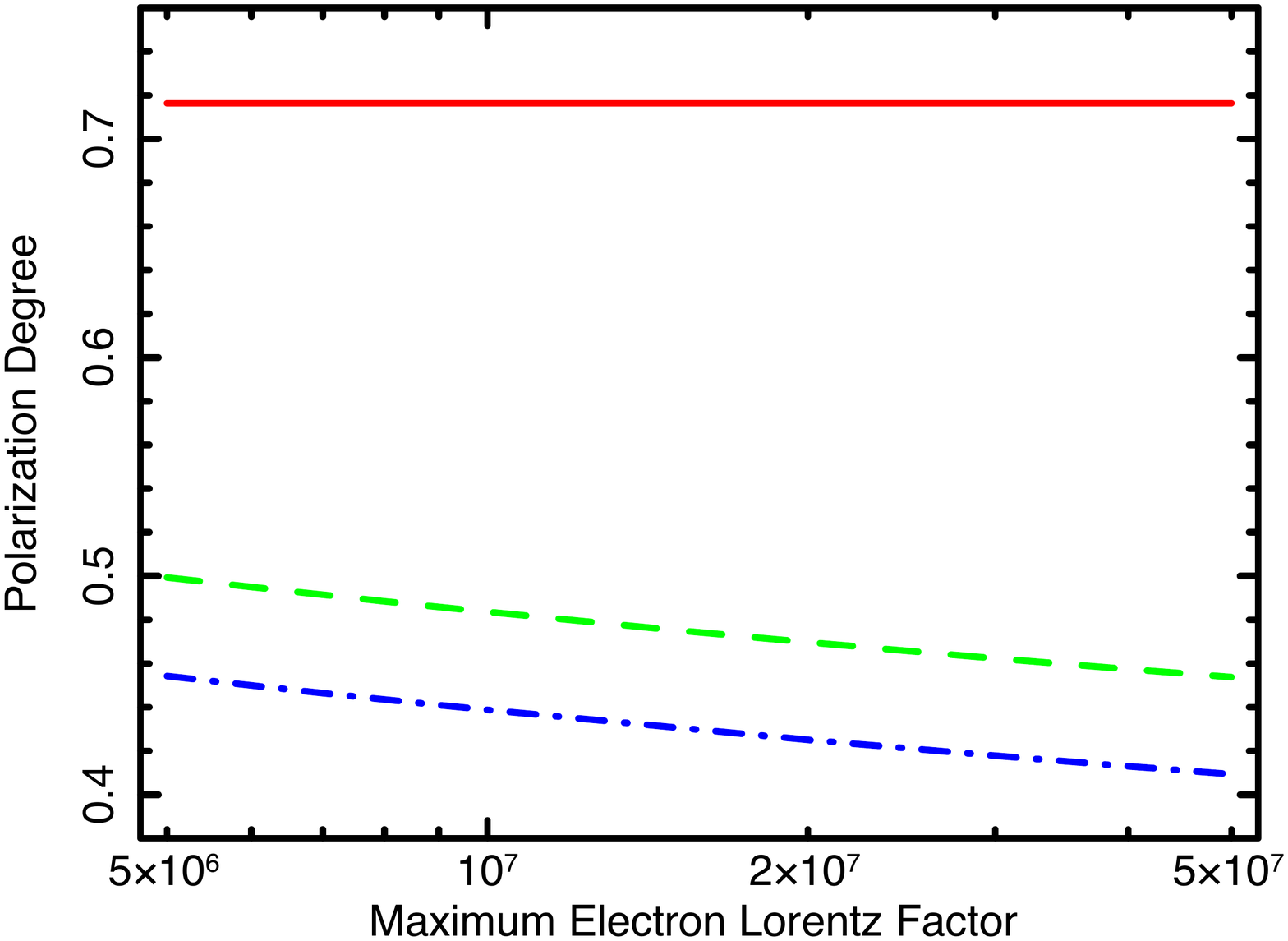}
\caption{Left panel: polarization degree as a function of magnetic field in the case of $1<\gamma<10^7$ at 5 MeV. Right panel: polarization degree as a function of the maximum electron Lorentz factor $\gamma_{max}$ in the case of $B=1.0\times 10^6$ G at 5 MeV. The results are presented by the solid, dashed, and dot-dashed lines when we take $\gamma_{th}=5.5\times 10^{4},~2.0\times 10^4,~\rm{and}~3.0\times 10^{3}$, respectively. The electron temperature is given as $T_e=1.0\times 10^{13}$ K in the calculations.
\label{fig4}}
\end{figure}

\end{document}